\documentclass[twocolumn,prl,aps,tightenlines]{revtex4}

\usepackage{graphicx}
\usepackage{amsmath}
\usepackage{amssymb}
\usepackage{color}
\usepackage[utf8]{inputenc}
\usepackage[normalem]{ulem}
\usepackage{hyperref}

\def\bea{\begin{eqnarray}}
\def\eea{\end{eqnarray}}
\def\ba{\begin{eqnarray}}
\def\ea{\end{eqnarray}}
\def\be{\begin{equation}}
\def\ee{\end{equation}}
\def\beq{\begin{equation}}
\def\eeq{\end{equation}}

\newcommand{\Vsa}{V_{s, \,{\rm asym}}}

\begin{document}

\title{Thermalizing  Sterile Neutrino Dark Matter}

\author{\textbf{Rasmus S. L. Hansen}\hspace*{1mm}}\email{rasmus@mpi-hd.mpg.de}
\author{\textbf{Stefan Vogl}\hspace*{1mm}}\email{stefan.vogl@mpi-hd.mpg.de}
\affiliation{Max-Planck-Institut f\"ur Kernphysik,
Saupfercheckweg 1,
69117 Heidelberg,
Germany}

\begin{abstract}
Sterile neutrinos produced through oscillations are a well motivated dark matter candidate, but recent constraints from observations have ruled out most of the parameter space.
We analyze the impact of new interactions on the evolution of keV sterile neutrino dark matter in the early Universe. Based on general considerations we find a mechanism which thermalizes the sterile neutrinos after an initial production by oscillations. The thermalization of sterile neutrinos is accompanied by dark entropy production which increases the yield of dark matter and leads to a lower characteristic momentum. This resolves the growing tensions with structure formation and X-ray observations and even revives simple non-resonant production as a viable way to produce sterile neutrino dark matter. We investigate the parameters required for the realization of the thermalization mechanism in a representative model and find that a simple estimate based on energy and entropy conservation describes the mechanism well.  
\end{abstract}


\maketitle

{\bf Introduction}

Sterile neutrinos ($\nu_s$) are characterized by being Standard Model (SM) singlets, and in order to constitute a good dark matter candidate, they cannot be produced by a WIMP-like freeze-out mechanism. Instead, keV sterile neutrino dark matter can be produced non-thermally by oscillations of SM neutrinos  \cite{Dodelson:1993je}, the so-called Dodelson-Widrow mechanism (DW), which can be resonantly enhanced in the presence of a substantial lepton asymmetry \cite{Shi:1998km,Abazajian:2001nj}. 

Despite its small interaction with the SM, sterile neutrino dark matter leads to intriguing observable effects. On the one hand,  sterile neutrinos with masses in the keV range are warm dark matter candidates and leave an imprint on the large scale structure of the Universe which can be resolved by observations of the Lyman-alpha forests \cite{Viel:2013apy,Baur:2015jsy} or subhalo counts \cite{Horiuchi:2015qri}. On the other hand, the tiny mixing with active SM neutrinos required by production through oscillations induces the decay of a sterile to an active neutrino in association with a photon. The rate and the photon energy of this process are in reach of current \cite{Boyarsky:2007ge, Horiuchi:2013noa, Loewenstein:2008yi, Neronov:2016wdd, Perez:2016tcq, Yuksel:2007xh} and future~\cite{Neronov:2015kca} X-ray telescopes. 
Recently, the analysis of data taken by X-ray satellites found an unassociated X-ray line at 3.5 keV with an intensity which fits the expectation for decaying sterile neutrinos \cite{Bulbul:2014sua,Boyarsky:2014jta}. 
However, it seems that the entire parameter space preferred by production through oscillations is increasingly at odds with structure formation  when combined with X-ray bounds~\cite{Merle:2015vzu,Schneider:2016uqi,Cherry:2017dwu}. This is, in particular, also true for the tentative 3.5 keV-line.

There are a number of alternative production mechanisms for sterile neutrinos~\cite{Kusenko:2006rh,Asaka:2006ek,Asaka:2006nq,Merle:2013wta,Merle:2015oja,Berlin:2016bdv}, which open new windows for keV sterile neutrinos as dark matter. We will pursue an orthogonal approach 
and investigate whether a modified evolution of the sterile neutrinos after  their initial 
production can resolve the mounting tensions between theoretical expectations and observations.  
We find that interactions among the sterile neutrinos themselves and with other particles in a dark sector can alter their cosmological history substantially through thermalization and associated dark entropy production which lead to a cooler and more abundant population of sterile neutrinos.

The structure of this Letter is as follows. First, we briefly sketch the general mechanism for thermalizing sterile neutrinos produced by oscillations and present a simple analytic estimate for the expected temperature and number density. Next, we introduce a simplified model which will allow a more quantitative  discussion. After briefly reviewing the production of sterile neutrinos and the bounds on the model parameters that can be derived from it, we present a numerical calculation of the thermalization process. Finally, we discuss the impact on the allowed parameter space for keV sterile neutrinos as dark matter and present our conclusions.

{\bf General mechanism}

Consider a sterile neutrino which interacts with a new boson $X$ with a mass that fulfills $m_{\nu_s} \ll m_X \ll T_{\nu_s, {\rm production}}$ and assume that $X$ has additional couplings which enable number changing processes.
With these ingredients the  mechanism for producing and thermalizing sterile neutrinos proceeds as follows:
\begin{enumerate}
\item[I] $\mathbf{T\sim100MeV}$: $\nu_s$ and $\bar\nu_s$ are produced out-of-chemical equilibrium via oscillations.
\item[II] $\mathbf{T\sim100-10MeV}$: $\nu_s$ and $\bar\nu_s$  interact and produce $X$. The $X$ reaches chemical equilibrium via a rapid number changing processes.
\item[III] $\mathbf{T\sim10-1MeV}$: Once a sufficient abundance of $X$ has been built up, the production of $\nu_s$ and $\bar\nu_s$ from $X$ becomes efficient and $\nu_s$ and $\bar\nu_s$ are also driven towards chemical equilibrium.
\item[IV] $\mathbf{T\sim1MeV}$: The $X$ particles becomes non-relativistic and annihilate or decay to $\nu_s\bar\nu_s$.
\end{enumerate}
The thermalization in stage III is accompanied by entropy production in the dark sector. In contrast to entropy production for SM particles~\cite{Asaka:2006ek}, which dilutes the sterile neutrinos, this dark entropy production enhances the number density of $\nu_s$ and cools the distribution.

{\bf Simple estimate}

The energy density of a relativistic species scales as $\rho \propto a^{-4}$, where $a$ is the scale factor. Therefore, energy conservation in the dark sector implies that the energy density after the sterile neutrinos  have reached equilibrium but before $X$ annihilates, $\rho_{s,\textrm{eq}}$, has to fulfill the equality $\rho_{\nu_s,\textrm{initial}}(a_i) a_i^4 = \rho_{s,\textrm{eq}}(a_X) a_X^4$, where $a_i$ is taken before equilibration and $a_X$ after.

In the simplest scenario, the distribution function of sterile neutrinos produced by the DW mechanism can be approximated as $f_{\nu_s} \simeq \frac{1}{\Lambda} f_{\nu_a}$~\cite{Dodelson:1993je}, where $\Lambda\gg 1$ is a suppression factor parametrizing the underabundance of $\nu_s$ and $f_{\nu_a}$ is a Fermi-Dirac distribution function describing a thermalized species with the temperature of the SM bath. Combing this estimate with energy conservation, we find
\begin{equation}
  T_X= \left(\frac{2}{2+\frac{8}{7}g_{X}}\right)^{1/4}\Lambda^{-1/4} T_\gamma ,
\end{equation}
where $g_X$ denotes the number of degrees of freedom in $X$ and we have used that $T_\gamma \propto a^{-1}$.
Afterwards, $X$ becomes non-relativistic and number changing self-annihilations combined with decays transfer its entropy to the sterile neutrinos: $s(a_X)a_X^3 = s(a_{f})a_{f}^3$, where $a_f$ is taken after annihilation. This heats the $\nu_s$ and we find the final temperature
\begin{equation}
  T_{f} 
  = \left(1 + \frac{4}{7} g_X\right)^{1/12} \Lambda^{-1/4} T_\gamma .
\end{equation}
The comoving number density after $X$ annihilation and decay in terms of the initial number density is $n_{s,f}/s_{\rm SM}= (1+\frac{4}{7} g_X)^{1/4} \Lambda^{1/4} n_{s,i}/s_{\rm SM}$. A sterile neutrino with a thermal spectrum is therefore more abundant than in the standard DW case and, in addition, it is colder for reasonable values of $\Lambda$ and $g_X$.
 
Realistic initial distribution functions for sterile neutrinos are not quite as simple as $\frac{1}{\Lambda} f_{\nu_a}$, and the proper initial energy density of the system should be used instead. In addition, the entropy conservation argument needs to take into account that sterile neutrinos from resonant production possess an asymmetry. However, for all realistic asymmetries the effect of this on the final yield and temperature is smaller than $2\%$ and can safely be neglected.

{\bf Toy Model}

In order to assess the impact of new interactions in a dark sector more quantitatively, we introduce a toy model with a new scalar boson $\varphi$ interacting with sterile neutrinos $\nu_s$. 
We consider a generic Lagrangian for $\varphi$, which is given by
\begin{align}
\mathcal{L}_\varphi= \frac{1}{2}\partial^\mu \varphi \partial_\mu \varphi - \frac{1}{2} m_\varphi^2 \varphi^2 - \frac{\lambda}{4} \varphi^4
\end{align} 
and contains a self-interaction with strength $\lambda$.
In addition we introduce an interaction between $\varphi$ and the sterile neutrino 
which is described by
\begin{align}
\mathcal{L}_{\rm int} = y \bar{\nu}_s \nu_s \varphi ,
\end{align}
 where $y$ denotes a Yukawa coupling. We assume that $\varphi$ has no interactions with  SM particles but in principle a Higgs portal interaction with the new scalar $\varphi$ is also possible. Such an interaction could provide an alternative connection between the SM bath and the sterile sector, see \cite{Heikinheimo:2016yds} for a related discussion. Together these  interactions allow two processes which modify the evolution of sterile neutrinos in the early Universe:
a) the decay/inverse decay $\varphi \leftrightarrow \nu_s \bar{\nu}_s $ with a width given by
\begin{align}
\Gamma_\varphi \approx \frac{1}{4 \pi} y^2 m_\varphi \;,
\end{align}
and b) number changing processes such as $2 \varphi \leftrightarrow 4 \varphi$.
In the non-relativistic limit, i.e. in the situation when the rate is lowest, the thermally averaged cross section can be inferred from the inverse process~\cite{Bernal:2015xba} and reads
 \begin{align}
\sigma v_{2 \rightarrow 4 } \approx  \frac{27 \sqrt{3} }{64\pi^4} \frac{\lambda^4 T^3}{m_\varphi^5} \exp \left(-\frac{2 m_\varphi}{T} \right)\;.
\label{eq:2to4}
\end{align}
This model is  chosen for  illustration only  and any model which allows efficient number changing processes will lead to  similar results.    

{\bf Production of sterile neutrinos}

We assume that oscillations between one active neutrino $\nu_a$ (we consider $a=\mu$ without loss of generality) and the sterile neutrino are relevant for the production of $\nu_s$. 
The mixing angle between $\nu_s$ and $\nu_a$ is $\theta$, and the mass squared difference is approximately given by the sterile mass squared: $\Delta m^2 \approx m_s^2$.

The production of sterile neutrinos can to good approximation be described by the Boltzmann equation~\cite{DiBari:1999ha,Foot:1996qc}
\begin{widetext}
\begin{equation}
  \frac{\partial}{\partial t} f_{\nu_s}(p,t) - H p \frac{\partial}{\partial p} f_{\nu_s}(p,t) \approx \frac{1}{4} \frac{\Gamma_a(p) \Delta^2(p) \sin^22\theta}{\Delta^2(p) \sin^22\theta + D^2(p) + [\Delta(p) \cos2\theta - V_T(p) - V_L(p)]^2} [f_{\nu_a}(p,t) - f_{\nu_s}(p,t)] .
\end{equation}
\end{widetext}
Here $f_{\nu_s}(p,t)$ and $f_{\nu_a}(p,t)$ are the distributions of sterile and active neutrinos as a function of momentum $p$ and time $t$. $H$ is the Hubble constant, and $\Delta(p) = \Delta m^2/2p$. At temperatures $T\lesssim 100$MeV, the collision rates are approximately
\begin{equation}
  \Gamma_a(p) = 
  \begin{cases}
    1.27 G_F^2 p T^4, & a=e,\\
    0.92 G_F^2 p T^4, & a=\mu,\tau \;,
  \end{cases}
\end{equation}
where $G_F$ is the Fermi constant.
The damping term is $D(p) = \Gamma_a(p)/2$. The potentials are approximately given by 
\begin{align}
  V_T(p) &= \quad-\frac{8 \sqrt{2} G_F p}{3 m_Z^2} (\rho_{\nu_a} + \rho_{\bar\nu_a}) 
  -\frac{8 \sqrt{2} G_F p}{3 m_W^2} (\rho_{a} + \rho_{\bar a}),\\
  V_L(p) &= \sqrt{2} G_F \left[ 2 \Delta n_{\nu_a} + \sum_{b\neq a} \Delta n_{\nu_b}  + \Delta n_a- \frac{n_n}{2}\right]
\end{align}
where $n_k$ and $\rho_k$ are number- and energy densities for the particle $k$ ($k=a$, charged lepton; $k=n$, neutron), while $\Delta n_k = n_k-n_{\bar k}$. For DW production the asymmetries are assumed to be negligible, but in the resonant case, a large lepton asymmetry $L_a = (n_{\nu_a} - n_{\bar\nu_a})/s$ is present. 

We use the public code `sterile-dm'~\cite{Venumadhav:2015pla} to solve the Boltzmann equation. It includes additional corrections in the treatment of the collision rates, the potential and the lepton asymmetry.

The introduction of an interaction between $\nu_s$ and $\varphi$ could change the oscillational production of $\nu_s$ in two important ways. First, a new potential is introduced if an asymmetry is present in sterile neutrinos~\cite{Dasgupta:2013zpn}
\begin{equation}
  \Vsa{}(p) = \frac{y^2}{2m_\varphi^2} \Delta n_{\nu_s}.
\end{equation}
Second, the momentum-integrated collision rate is modified due to inverse decays. After equilibration, the rate reads
\begin{equation}
  \Gamma_{\nu_s} = \frac{m_\varphi^2 \Gamma_\varphi T_\varphi}{2 \pi^2} \frac{K_1\left(\tfrac{m_\varphi}{T_\varphi}\right)}{n_{\nu_{s}}},
\end{equation}
 where $K_1$ is the modified Bessel function of the second kind, and Boltzmann statistics are assumed.

We assume in our calculations that the production of sterile neutrinos through oscillations is not affected by $\varphi$, and that the presence of $\varphi$ only affects $\nu_s$ through the thermalization mechanism. This assumption puts some bounds on the coupling constants that can be allowed.

If $\nu_s$ is produced resonantly, we require that $V_L > 10 \,\Vsa{}$ resulting in the bound
\begin{equation}
  y < \sqrt{\frac{2\sqrt{2} G_F m_\varphi^2 L s}{5 \Delta n_{\nu_s}}  } .
\end{equation}
The asymmetries $L$ and $\Delta n_{\nu_s}$ are non-trivial functions of temperature, but the strongest constraint arises when both have reached their final values. Using a grid in $m_{\nu_s}$ and $L$, we find that the strongest limits at $m_{\nu_s}=3,7$, and $50$ keV and $m_\varphi=0.1,0.1$, and $0.3$ MeV are
\begin{equation*}
  y<2\times 10^{-8}, \;\;\, y<1\times10^{-7} \;, \;\, \textrm{and} \;\;\,  y<2\times 10^{-6} .
\end{equation*}
The bound is strongest when $L$ is reduced significantly due to $\nu_s$ production and becomes weaker for high values of $L$. For low values of $L$ it becomes insignificant as DW production takes over.

At temperatures $T \sim m_\varphi$, the inverse decay rate becomes maximal, and can potentially lead to additional production of $\nu_s$ through oscillations at a rate $\Gamma_{\varphi, {\rm ID}} = \frac{1}{4} \sin^2 2\theta \Gamma_{\nu_s}$. 
The additional production must be compared to the production rate at high temperature due to the DW mechanism, $\Gamma_{\rm DW}$.
The condition 
$\Gamma_{\rm DW}(T_{\gamma,\rm DW}) / H(T_{\gamma,\rm DW}) > 10 \, \Gamma_{\varphi,{\rm ID}}(T_{\varphi,{\rm ID}}) / H(T_{\gamma,{\rm ID}})$
results in the limit
\begin{equation}
  y < 1.6 \times 10^{-8} \left(\frac{g_*(T_{\gamma,{\rm ID}})}{g_*(T_{\gamma,\rm DW})}\right)^{1/4} \sqrt{\frac{m_{\nu_s}}{\rm keV} \frac{m_\varphi}{\rm MeV}} \Lambda^{1/4},
\end{equation}
where $T_{\gamma/\varphi, {\rm ID}/{\rm DW}}$ refers to the photon-/$\varphi$-temperature of maximal production from $\Gamma_{\varphi,{\rm ID}}/ \Gamma_{\rm DW}$.
For the masses $m_{\nu_s} = 3, 7$, and $50$ keV, the limits are
\begin{equation*}
  y < 3\times 10^{-8}, \;\;\, y < 7\times10^{-8} \;,\;\, \textrm{and} \;\;\, y < 6\times10^{-7} ,
\end{equation*}
when $m_\varphi = 0.1, 0.1$ and $0.3$ MeV and $\Lambda$ is determined such that the thermalized sterile neutrinos give the correct dark matter abundance.
The bound weakens when resonant production dominates over non-resonant production since $\Gamma_{\varphi,{\rm ID}}$ decreases with $\sin^2 2\theta$ while the resonance keeps the production at higher temperature efficient.

{\bf Numerical calculation of thermalization}

The cosmological evolution of the different species in the dark sector can be described by a system of coupled Boltzmann equations, which track the distributions of $\nu_s, \bar{\nu}_s $ and $\varphi$.  We reduce the complexity of this problem by using the following simplifying assumptions. Quantum statistic factors are neglected and we use Boltzmann statistics for all involved particle species. All particles are  taken to be in local thermodynamic equilibrium, i.e. we describe their distribution functions by temperatures $T_i$ and  chemical potentials $\mu_i$.   The sterile neutrinos are lighter than $\varphi$ and can be treated as massless, whereas the $m_\varphi$-dependence is taken into account.
Finally,  we assume that the number changing processes of the $\varphi$s are rapid, i.e $\mu_\varphi=0$ .  

Now three energy densities ($\rho_\varphi$, $\rho_{\nu_s}$, and $\rho_{\bar{\nu}_s}$)  and two number densities ($n_{\nu_s}$ and $n_{\bar{\nu}_s}$) characterize the system.  The evolution of the densities is given by the integrated Boltzmann equations
\begin{equation}
\begin{aligned}
\dot{\rho}_{\varphi} + C H \rho_{\varphi} &=  \Gamma_{\rho_{\nu_s}}  \rho_{\nu_s} + \Gamma_{\rho_{\bar{\nu}_s}}  \rho_{\bar{\nu}_s} - \Gamma_{\rho_\varphi}  \rho_\varphi \\
\dot{\rho}_{\nu_s} + 4 H \rho_{\nu_s} &= \Gamma_{\rho_\varphi}\rho_\varphi/2 - \Gamma_{\rho_{\nu_s}}  \rho_{\nu_s} \\
\dot{\rho}_{\bar{\nu}_s} + 4 H \rho_{\bar{\nu}_s} &= \Gamma_{\rho_\varphi}  \rho_\varphi /2 - \Gamma_{\rho_{\bar{\nu}_s}}  \rho_{\bar{\nu}_s} \\
\dot{n}_{\nu_s} + 3 H n_{\nu_s}&= \Gamma_{n_\varphi} n_\varphi - \Gamma_{n_{\nu_s}} n_{\nu_s} \\
\dot{n}_{\bar{\nu}_s} + 3 H n_{\bar{\nu}_s}&= \Gamma_{n_\varphi} n_\varphi - \Gamma_{n_{\bar{\nu}_s}} n_{\bar{\nu}_s}\  \;,
\end{aligned}
\label{eq:thermalization}
\end{equation}
where 
$C = \frac{1}{2 \pi^2 \rho_\varphi}\int d p (p^4 E^{-1} +3 p^2 E) f_\varphi(p,t)  $ accounts for the transition of $\varphi$ from the relativistic to the non-relativistic regime.
 In the high energy limit the interaction rates for $\varphi$ decay are given by
\begin{align}
\Gamma_{n_\varphi}=\frac{3}{2} \Gamma_{\rho_\varphi}=\frac{1}{2}\frac{m_\varphi}{T_\varphi}\Gamma_\varphi ,
\end{align}
while  the inverse decay rates of $\nu_s$ read
\begin{align}
\Gamma_{n_{\nu_s}}=3 \Gamma_{\rho_{\nu_s}}=\frac{1}{2} \frac{m_\varphi T_{\bar{\nu}_s}}{T_{\nu_s}^2} \exp\left[ \frac{\mu_{\bar{\nu}_s}}{T_{\bar{\nu}_s}} \right] \Gamma_\varphi\,.
\end{align} 
The rates for $\bar{\nu}_s$ are analogous to those for $\nu_s$ and can be obtained by exchanging the temperatures and the chemical potentials appropriately. The rates with the full $m_\varphi$-dependence are lengthy and we do not report them here, but  they are implemented in our numerical calculations.

\begin{figure}[tpb]
  \centering
  \includegraphics[width=\columnwidth]{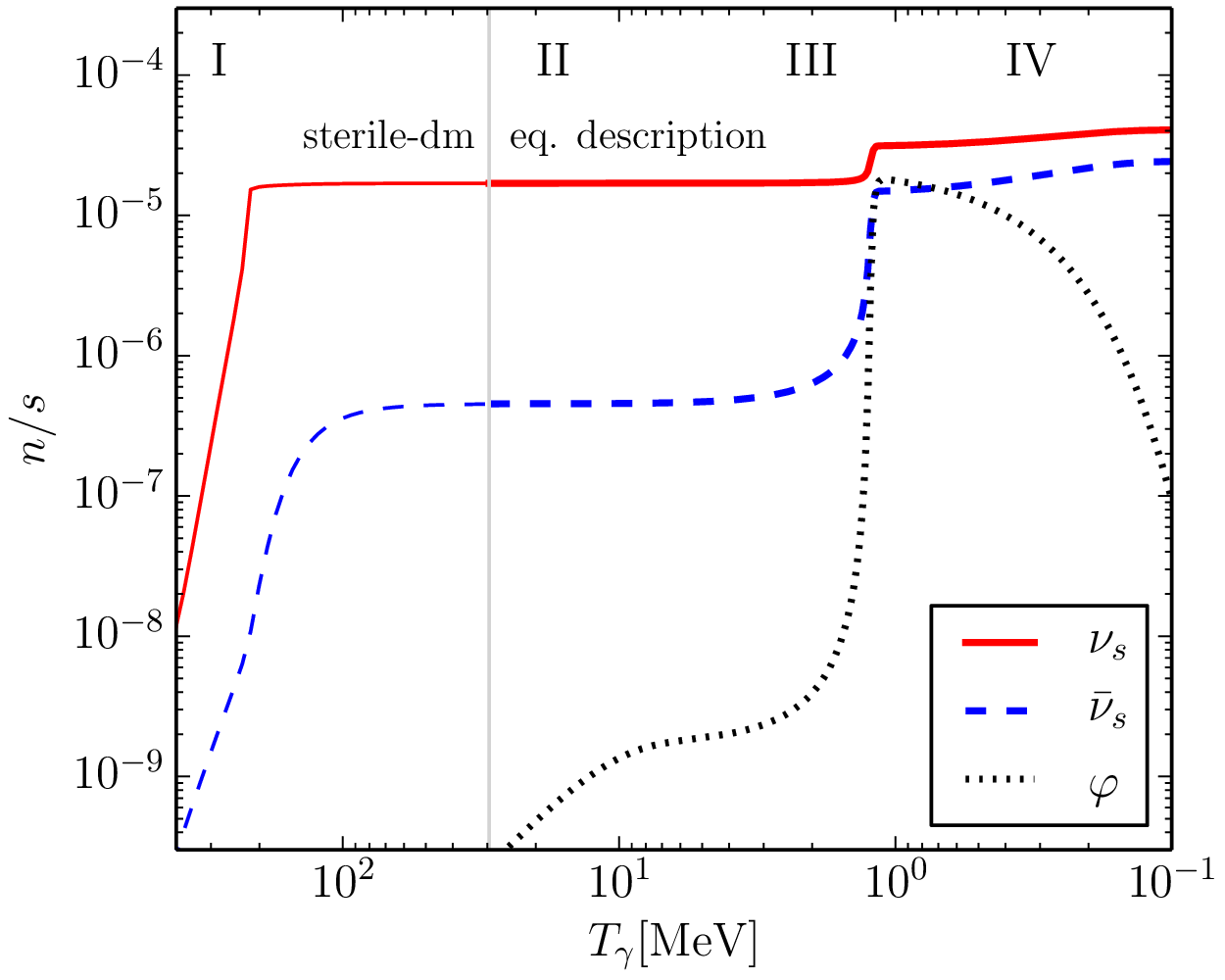}
  \caption{{\bf Thermalization mechanism.}  Abundances  of the dark sector species as a function of $T_\gamma$ for $m_{\nu_s}= 7$ keV, $m_\varphi= 0.1$ MeV, $n_{\bar{\nu}_s}/n_{\nu_s}= 3 \times 10^{-2}$, and $y= 7\times 10^{-9}$.  The epochs, I-IV, correspond to the stages of the general mechanism. We use the code `sterile-dm' at temperature to the left of the gray line. For temperatures to the right of the line, the local thermal equilibrium description in Eq.~(\ref{eq:thermalization}) is used. }
  \label{fig:thermalise}
\end{figure}

As can be seen in Fig.~\ref{fig:thermalise}, the production proceeds in the four stages we described for the general mechanism.
First, an initial abundance of $\nu_s$ is built up by DW or resonant production (I). Then the inverse decay $\nu_s \bar{\nu}_{s} \rightarrow \varphi$ produces a bath of $\varphi$ particles (II). After thermalizing with themselves, the decay rate of $\varphi$ starts to have an impact on the $\bar{\nu}_s$ abundance leading to an increase of its number density and a drop in temperature (III). This drives up the inverse decay rate of $\nu_s$ and efficient thermal contact between $\nu_s$, $\bar{\nu}_s$, and $\varphi$ is established. Finally, $\varphi$ becomes non-relativistic, and number changing $4 \rightarrow 2$ annihilations heat the system thus transferring the entropy in $\varphi$ to the $\nu_s$-bath (IV).

Taking again the masses $m_{\nu_s} = 3, 7$, and $50$ keV, and $m_\varphi = 0.1, 0.1$, and $0.3$ MeV as representative values, we find that
\begin{equation*}
  y > 6\times 10^{-9}, \;\;\, y > 6\times10^{-9} \;,\;\,\textrm{and} \;\;\, y > 3\times10^{-8} 
\end{equation*}
allow a successful thermalization before $T_\gamma= 1$ MeV with only a marginal dependence on the asymmetry. 
The final number densities and temperatures agree excellently with the estimate presented previously provided the use of Boltzmann statistics is accounted for.

The value of $\lambda$ which controls the  strength of the number changing process $2 \varphi \leftrightarrow 4 \varphi$ should also be considered.
In order to avoid a population of hot $\nu_s$ from the decay of frozen-out $\varphi$s,  which might spoil the warm dark
 matter bound, we require that $2 \varphi \leftrightarrow 4\varphi$ 
remains faster than the Hubble rate until $n_\varphi < 10^{-3} n_{\nu_s}$. Taking the analytical estimate for the thermally averaged cross section in Eq.~(\ref{eq:2to4}), this leads to $\lambda=\mathcal{O}(0.1)$. 
 
For strongly asymmetric initial conditions, $n_{\bar\nu_s}/n_{\nu_s}< 10^{-4}$ for $m_s>7$ keV (corresponding to $L\sim 10^{-3}$--$10^{-4}$), we have not been able to solve the system of equations in Eq.~(\ref{eq:thermalization}) 
numerically. 
The computation breaks down due to the very sudden thermalization. Before equilibrium between $\nu_s$ and $\varphi$ is reached, $T_\varphi$ overshoots $T_{\nu_s}$, and this overshoot becomes stronger as the asymmetry becomes larger leading to a numerical instability. 
At the same time the value of $\lambda$ required to ensure efficient number changing processes throughout the evolution of the system grows and it approaches $\lambda \sim 1$ when our numerical solution fails. Therefore, the simplifying assumption that $\mu_\varphi = 0$ is harder to satisfy for large asymmetries. 
Although the momentum averaged description becomes questionable in this regime, there is no reason to expect that the thermalization mechanism does not work. However, a full momentum dependent description is needed in this limit.

\begin{figure}[tbp]
  \centering
  \includegraphics[width=\columnwidth]{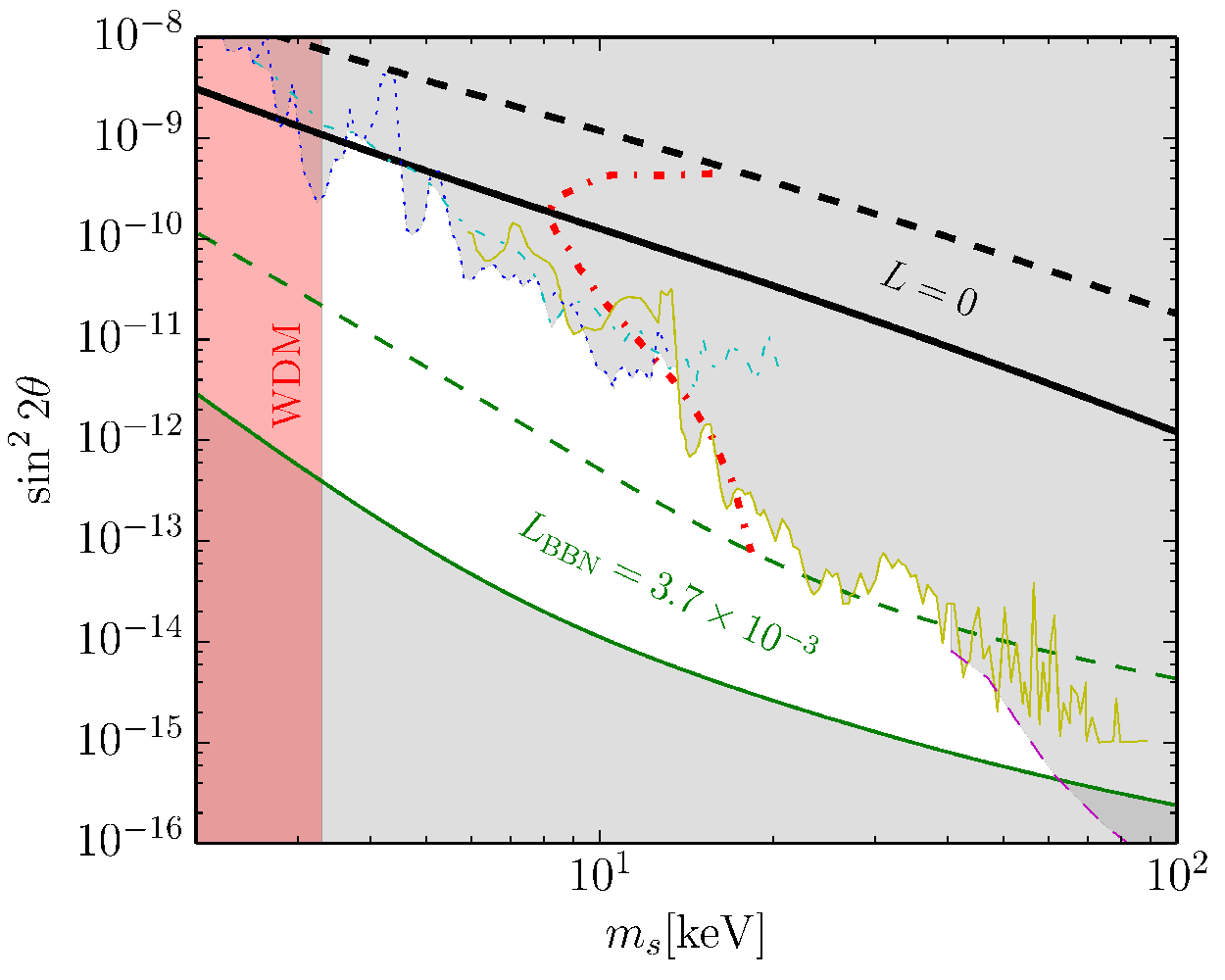}
  \caption{{\bf Allowed regions in mixing parameters.} 
The solid (dashed) thick black and thinner green lines give the correct dark matter abundance for a thermalized (not thermalized) $\nu_s$ in the DW case and the limit from big bang nucleosynthesis. The red shaded area (dot-dashed red line) gives the limit from the Lyman-$\alpha$ forest for a thermal~\cite{Viel:2013apy} (non-thermal~\cite{Schneider:2016uqi}) $\nu_s$. The upper gray region is excluded by X-ray observations from Chandra~\cite{Horiuchi:2013noa} (dotted blue line), Suzaku~\cite{Loewenstein:2008yi} (dot-dashed cyan line), NuStar~\cite{Perez:2016tcq} (yellow line), and Integral~\cite{Yuksel:2007xh} (dashed magenta line).}
  \label{fig:res}
\end{figure}

{\bf Results and conclusions}

The thermalization process increases the number density of sterile neutrinos through dark entropy production. Therefore, the initial abundance produced by oscillations of SM neutrinos is lower than in the standard case. A lower initial yield points towards smaller values of the mixing parameter $\sin^2 2 \theta$. As can be seen in Fig.~\ref{fig:res}, thermalization shifts the expected parameter space down by about one order of magnitude and reduces the tensions with X-ray observations substantially. In particular the DW production mechanism, which is already excluded by X-ray observations for non-interacting sterile neutrinos, remains viable in the vicinity of $m_{\nu_s}=4 $ keV. 
It should be kept in mind that uncertainties in the predictions from DW production and the X-ray limits could modify this conclusion.
Since DW production happens during the QCD phase transition, the equation of state and hadronic scattering rates are hard to determine and the sterile neutrino yield could be enhanced or suppressed by a factor of 2~\cite{Asaka:2006nq}. 
In addition,  X-ray bounds are subject to astrophysical uncertainties. For instance, the mass determination of the dwarf galaxy Ursa Minor, i.e. the target of~\cite{Loewenstein:2008yi}, is known to be affected by systematic uncertainties.  As a result, the dark matter mass and consequently the X-ray flux could be a factor of 2 higher or lower~\cite{Bonnivard:2015xpq}. The effect of similar uncertainties on other limits is harder to estimate since the structure of these objects is more complicated. When the uncertainties in the production and the dwarf limits are taken into account, DW remains viable for $m_s ~< 5$ keV.
If the uncertainty of other X-ray searches should be similar, the viable mass range reaches up to 6 keV. 

In addition, thermalization decreases the average momentum $\langle p \rangle$ of $\nu_s$, i.e. cools the sterile neutrinos. 
Since the final population of $\nu_s$ is in equilibrium, the momentum distribution is thermal and astrophysical bounds on thermal warm dark matter apply directly. 
This is an important difference compared with sterile neutrinos produced by the DW mechanism or resonant production. The limit on warm dark matter from the Lyman-alpha forest requires $m_{\nu_s}\geq 3.3$  keV~\cite{Viel:2013apy}. For comparison we also show the limits on sterile neutrinos with a momentum distribution expected from production purely through oscillations from \cite{Schneider:2016uqi}. As can be seen, these limits are much more stringent and exclude $m_{\nu_s} \lesssim 16 $ keV for the  DW mechanism.
Bounds from Big Bang Nucleosynthesis put an upper limit on $L$~\cite{Castorina:2012md}, which translates into a lower limit on $\sin^22\theta$, and this limit is also relaxed by the thermalization mechanism.
The combined effect on the momentum distribution and the number density relaxes the constraints substantially, and we find a sizable region of parameter space where thermalized sterile neutrinos can account for the dark matter abundance in the Universe.  

The simple mechanism discussed here does not predict new signatures for sterile neutrino dark matter. However, the relation between the impact on structure formation and the decay rates into X-ray lines is modified compared to the prediction from production by oscillation. Therefore, a detection in both channels could provide evidence for thermalized sterile neutrinos.

Finally, we would like to stress that new interactions of
sterile neutrinos could have consequences which go beyond
mere thermalization. If the interactions are strong enough
they could have a direct impact on the production of sterile
neutrinos through oscillations by enhancing the interaction
rates and by modifying the dispersion relations. In such a
scenario a further boost of the sterile neutrinos abundance
can be expected, and there is a potential that new regions of
parameter space open up.

{\bf Acknowledgments}

The authors would like to thank Kai Schmitz and Christian Vogl for useful discussions and Maximilian Totzauer and Alexander Merle for valuable comments. RSLH is funded by the Alexander von Humboldt Foundation.

\bibliography{note}

\begin{thebibliography}{32}
\expandafter\ifx\csname natexlab\endcsname\relax\def\natexlab#1{#1}\fi
\expandafter\ifx\csname bibnamefont\endcsname\relax
  \def\bibnamefont#1{#1}\fi
\expandafter\ifx\csname bibfnamefont\endcsname\relax
  \def\bibfnamefont#1{#1}\fi
\expandafter\ifx\csname citenamefont\endcsname\relax
  \def\citenamefont#1{#1}\fi
\expandafter\ifx\csname url\endcsname\relax
  \def\url#1{\texttt{#1}}\fi
\expandafter\ifx\csname urlprefix\endcsname\relax\def\urlprefix{URL }\fi
\providecommand{\bibinfo}[2]{#2}
\providecommand{\eprint}[2][]{\url{#2}}

\bibitem[{\citenamefont{Dodelson and Widrow}(1994)}]{Dodelson:1993je}
\bibinfo{author}{\bibfnamefont{S.}~\bibnamefont{Dodelson}} \bibnamefont{and}
  \bibinfo{author}{\bibfnamefont{L.~M.} \bibnamefont{Widrow}},
  \bibinfo{journal}{Phys. Rev. Lett.} \textbf{\bibinfo{volume}{72}},
  \bibinfo{pages}{17} (\bibinfo{year}{1994}), \eprint{hep-ph/9303287}.

\bibitem[{\citenamefont{Shi and Fuller}(1999)}]{Shi:1998km}
\bibinfo{author}{\bibfnamefont{X.-D.} \bibnamefont{Shi}} \bibnamefont{and}
  \bibinfo{author}{\bibfnamefont{G.~M.} \bibnamefont{Fuller}},
  \bibinfo{journal}{Phys. Rev. Lett.} \textbf{\bibinfo{volume}{82}},
  \bibinfo{pages}{2832} (\bibinfo{year}{1999}), \eprint{astro-ph/9810076}.

\bibitem[{\citenamefont{Abazajian et~al.}(2001)\citenamefont{Abazajian, Fuller,
  and Patel}}]{Abazajian:2001nj}
\bibinfo{author}{\bibfnamefont{K.}~\bibnamefont{Abazajian}},
  \bibinfo{author}{\bibfnamefont{G.~M.} \bibnamefont{Fuller}},
  \bibnamefont{and} \bibinfo{author}{\bibfnamefont{M.}~\bibnamefont{Patel}},
  \bibinfo{journal}{Phys. Rev.} \textbf{\bibinfo{volume}{D64}},
  \bibinfo{pages}{023501} (\bibinfo{year}{2001}), \eprint{astro-ph/0101524}.

\bibitem[{\citenamefont{Viel et~al.}(2013)\citenamefont{Viel, Becker, Bolton,
  and Haehnelt}}]{Viel:2013apy}
\bibinfo{author}{\bibfnamefont{M.}~\bibnamefont{Viel}},
  \bibinfo{author}{\bibfnamefont{G.~D.} \bibnamefont{Becker}},
  \bibinfo{author}{\bibfnamefont{J.~S.} \bibnamefont{Bolton}},
  \bibnamefont{and} \bibinfo{author}{\bibfnamefont{M.~G.}
  \bibnamefont{Haehnelt}}, \bibinfo{journal}{Phys. Rev.}
  \textbf{\bibinfo{volume}{D88}}, \bibinfo{pages}{043502}
  (\bibinfo{year}{2013}), \eprint{1306.2314}.

\bibitem[{\citenamefont{Baur et~al.}(2016)\citenamefont{Baur,
  Palanque-Delabrouille, Yèche, Magneville, and Viel}}]{Baur:2015jsy}
\bibinfo{author}{\bibfnamefont{J.}~\bibnamefont{Baur}},
  \bibinfo{author}{\bibfnamefont{N.}~\bibnamefont{Palanque-Delabrouille}},
  \bibinfo{author}{\bibfnamefont{C.}~\bibnamefont{Yèche}},
  \bibinfo{author}{\bibfnamefont{C.}~\bibnamefont{Magneville}},
  \bibnamefont{and} \bibinfo{author}{\bibfnamefont{M.}~\bibnamefont{Viel}},
  \bibinfo{journal}{JCAP} \textbf{\bibinfo{volume}{1608}}, \bibinfo{pages}{012}
  (\bibinfo{year}{2016}), \eprint{1512.01981}.

\bibitem[{\citenamefont{Horiuchi et~al.}(2016)\citenamefont{Horiuchi, Bozek,
  Abazajian, Boylan-Kolchin, Bullock, Garrison-Kimmel, and
  Onorbe}}]{Horiuchi:2015qri}
\bibinfo{author}{\bibfnamefont{S.}~\bibnamefont{Horiuchi}},
  \bibinfo{author}{\bibfnamefont{B.}~\bibnamefont{Bozek}},
  \bibinfo{author}{\bibfnamefont{K.~N.} \bibnamefont{Abazajian}},
  \bibinfo{author}{\bibfnamefont{M.}~\bibnamefont{Boylan-Kolchin}},
  \bibinfo{author}{\bibfnamefont{J.~S.} \bibnamefont{Bullock}},
  \bibinfo{author}{\bibfnamefont{S.}~\bibnamefont{Garrison-Kimmel}},
  \bibnamefont{and} \bibinfo{author}{\bibfnamefont{J.}~\bibnamefont{Onorbe}},
  \bibinfo{journal}{Mon. Not. Roy. Astron. Soc.}
  \textbf{\bibinfo{volume}{456}}, \bibinfo{pages}{4346} (\bibinfo{year}{2016}),
  \eprint{1512.04548}.

\bibitem[{\citenamefont{Boyarsky et~al.}(2008)\citenamefont{Boyarsky, Malyshev,
  Neronov, and Ruchayskiy}}]{Boyarsky:2007ge}
\bibinfo{author}{\bibfnamefont{A.}~\bibnamefont{Boyarsky}},
  \bibinfo{author}{\bibfnamefont{D.}~\bibnamefont{Malyshev}},
  \bibinfo{author}{\bibfnamefont{A.}~\bibnamefont{Neronov}}, \bibnamefont{and}
  \bibinfo{author}{\bibfnamefont{O.}~\bibnamefont{Ruchayskiy}},
  \bibinfo{journal}{Mon. Not. Roy. Astron. Soc.}
  \textbf{\bibinfo{volume}{387}}, \bibinfo{pages}{1345} (\bibinfo{year}{2008}),
  \eprint{0710.4922}.

\bibitem[{\citenamefont{Horiuchi et~al.}(2014)\citenamefont{Horiuchi, Humphrey,
  Onorbe, Abazajian, Kaplinghat, and Garrison-Kimmel}}]{Horiuchi:2013noa}
\bibinfo{author}{\bibfnamefont{S.}~\bibnamefont{Horiuchi}},
  \bibinfo{author}{\bibfnamefont{P.~J.} \bibnamefont{Humphrey}},
  \bibinfo{author}{\bibfnamefont{J.}~\bibnamefont{Onorbe}},
  \bibinfo{author}{\bibfnamefont{K.~N.} \bibnamefont{Abazajian}},
  \bibinfo{author}{\bibfnamefont{M.}~\bibnamefont{Kaplinghat}},
  \bibnamefont{and}
  \bibinfo{author}{\bibfnamefont{S.}~\bibnamefont{Garrison-Kimmel}},
  \bibinfo{journal}{Phys. Rev.} \textbf{\bibinfo{volume}{D89}},
  \bibinfo{pages}{025017} (\bibinfo{year}{2014}), \eprint{1311.0282}.

\bibitem[{\citenamefont{Loewenstein et~al.}(2009)\citenamefont{Loewenstein,
  Kusenko, and Biermann}}]{Loewenstein:2008yi}
\bibinfo{author}{\bibfnamefont{M.}~\bibnamefont{Loewenstein}},
  \bibinfo{author}{\bibfnamefont{A.}~\bibnamefont{Kusenko}}, \bibnamefont{and}
  \bibinfo{author}{\bibfnamefont{P.~L.} \bibnamefont{Biermann}},
  \bibinfo{journal}{Astrophys. J.} \textbf{\bibinfo{volume}{700}},
  \bibinfo{pages}{426} (\bibinfo{year}{2009}), \eprint{0812.2710}.

\bibitem[{\citenamefont{Neronov et~al.}(2016)\citenamefont{Neronov, Malyshev,
  and Eckert}}]{Neronov:2016wdd}
\bibinfo{author}{\bibfnamefont{A.}~\bibnamefont{Neronov}},
  \bibinfo{author}{\bibfnamefont{D.}~\bibnamefont{Malyshev}}, \bibnamefont{and}
  \bibinfo{author}{\bibfnamefont{D.}~\bibnamefont{Eckert}},
  \bibinfo{journal}{Phys. Rev.} \textbf{\bibinfo{volume}{D94}},
  \bibinfo{pages}{123504} (\bibinfo{year}{2016}), \eprint{1607.07328}.

\bibitem[{\citenamefont{Perez et~al.}(2016)\citenamefont{Perez, Ng, Beacom,
  Hersh, Horiuchi, and Krivonos}}]{Perez:2016tcq}
\bibinfo{author}{\bibfnamefont{K.}~\bibnamefont{Perez}},
  \bibinfo{author}{\bibfnamefont{K.~C.~Y.} \bibnamefont{Ng}},
  \bibinfo{author}{\bibfnamefont{J.~F.} \bibnamefont{Beacom}},
  \bibinfo{author}{\bibfnamefont{C.}~\bibnamefont{Hersh}},
  \bibinfo{author}{\bibfnamefont{S.}~\bibnamefont{Horiuchi}}, \bibnamefont{and}
  \bibinfo{author}{\bibfnamefont{R.}~\bibnamefont{Krivonos}}
  (\bibinfo{year}{2016}), \eprint{1609.00667}.

\bibitem[{\citenamefont{Yuksel et~al.}(2008)\citenamefont{Yuksel, Beacom, and
  Watson}}]{Yuksel:2007xh}
\bibinfo{author}{\bibfnamefont{H.}~\bibnamefont{Yuksel}},
  \bibinfo{author}{\bibfnamefont{J.~F.} \bibnamefont{Beacom}},
  \bibnamefont{and} \bibinfo{author}{\bibfnamefont{C.~R.}
  \bibnamefont{Watson}}, \bibinfo{journal}{Phys. Rev. Lett.}
  \textbf{\bibinfo{volume}{101}}, \bibinfo{pages}{121301}
  (\bibinfo{year}{2008}), \eprint{0706.4084}.

\bibitem[{\citenamefont{Neronov and Malyshev}(2016)}]{Neronov:2015kca}
\bibinfo{author}{\bibfnamefont{A.}~\bibnamefont{Neronov}} \bibnamefont{and}
  \bibinfo{author}{\bibfnamefont{D.}~\bibnamefont{Malyshev}},
  \bibinfo{journal}{Phys. Rev.} \textbf{\bibinfo{volume}{D93}},
  \bibinfo{pages}{063518} (\bibinfo{year}{2016}), \eprint{1509.02758}.

\bibitem[{\citenamefont{Bulbul et~al.}(2014)\citenamefont{Bulbul, Markevitch,
  Foster, Smith, Loewenstein, and Randall}}]{Bulbul:2014sua}
\bibinfo{author}{\bibfnamefont{E.}~\bibnamefont{Bulbul}},
  \bibinfo{author}{\bibfnamefont{M.}~\bibnamefont{Markevitch}},
  \bibinfo{author}{\bibfnamefont{A.}~\bibnamefont{Foster}},
  \bibinfo{author}{\bibfnamefont{R.~K.} \bibnamefont{Smith}},
  \bibinfo{author}{\bibfnamefont{M.}~\bibnamefont{Loewenstein}},
  \bibnamefont{and} \bibinfo{author}{\bibfnamefont{S.~W.}
  \bibnamefont{Randall}}, \bibinfo{journal}{Astrophys. J.}
  \textbf{\bibinfo{volume}{789}}, \bibinfo{pages}{13} (\bibinfo{year}{2014}),
  \eprint{1402.2301}.

\bibitem[{\citenamefont{Boyarsky et~al.}(2014)\citenamefont{Boyarsky,
  Ruchayskiy, Iakubovskyi, and Franse}}]{Boyarsky:2014jta}
\bibinfo{author}{\bibfnamefont{A.}~\bibnamefont{Boyarsky}},
  \bibinfo{author}{\bibfnamefont{O.}~\bibnamefont{Ruchayskiy}},
  \bibinfo{author}{\bibfnamefont{D.}~\bibnamefont{Iakubovskyi}},
  \bibnamefont{and} \bibinfo{author}{\bibfnamefont{J.}~\bibnamefont{Franse}},
  \bibinfo{journal}{Phys. Rev. Lett.} \textbf{\bibinfo{volume}{113}},
  \bibinfo{pages}{251301} (\bibinfo{year}{2014}), \eprint{1402.4119}.

\bibitem[{\citenamefont{Merle et~al.}(2016)\citenamefont{Merle, Schneider, and
  Totzauer}}]{Merle:2015vzu}
\bibinfo{author}{\bibfnamefont{A.}~\bibnamefont{Merle}},
  \bibinfo{author}{\bibfnamefont{A.}~\bibnamefont{Schneider}},
  \bibnamefont{and} \bibinfo{author}{\bibfnamefont{M.}~\bibnamefont{Totzauer}},
  \bibinfo{journal}{JCAP} \textbf{\bibinfo{volume}{1604}}, \bibinfo{pages}{003}
  (\bibinfo{year}{2016}), \eprint{1512.05369}.

\bibitem[{\citenamefont{Schneider}(2016)}]{Schneider:2016uqi}
\bibinfo{author}{\bibfnamefont{A.}~\bibnamefont{Schneider}},
  \bibinfo{journal}{JCAP} \textbf{\bibinfo{volume}{1604}}, \bibinfo{pages}{059}
  (\bibinfo{year}{2016}), \eprint{1601.07553}.

\bibitem[{\citenamefont{Cherry and Horiuchi}(2017)}]{Cherry:2017dwu}
\bibinfo{author}{\bibfnamefont{J.~F.} \bibnamefont{Cherry}} \bibnamefont{and}
  \bibinfo{author}{\bibfnamefont{S.}~\bibnamefont{Horiuchi}},
  \bibinfo{journal}{Phys. Rev.} \textbf{\bibinfo{volume}{D95}},
  \bibinfo{pages}{083015} (\bibinfo{year}{2017}), \eprint{1701.07874}.

\bibitem[{\citenamefont{Kusenko}(2006)}]{Kusenko:2006rh}
\bibinfo{author}{\bibfnamefont{A.}~\bibnamefont{Kusenko}},
  \bibinfo{journal}{Phys. Rev. Lett.} \textbf{\bibinfo{volume}{97}},
  \bibinfo{pages}{241301} (\bibinfo{year}{2006}), \eprint{hep-ph/0609081}.

\bibitem[{\citenamefont{Asaka et~al.}(2006)\citenamefont{Asaka, Shaposhnikov,
  and Kusenko}}]{Asaka:2006ek}
\bibinfo{author}{\bibfnamefont{T.}~\bibnamefont{Asaka}},
  \bibinfo{author}{\bibfnamefont{M.}~\bibnamefont{Shaposhnikov}},
  \bibnamefont{and} \bibinfo{author}{\bibfnamefont{A.}~\bibnamefont{Kusenko}},
  \bibinfo{journal}{Phys. Lett.} \textbf{\bibinfo{volume}{B638}},
  \bibinfo{pages}{401} (\bibinfo{year}{2006}), \eprint{hep-ph/0602150}.

\bibitem[{\citenamefont{Asaka et~al.}(2007)\citenamefont{Asaka, Laine, and
  Shaposhnikov}}]{Asaka:2006nq}
\bibinfo{author}{\bibfnamefont{T.}~\bibnamefont{Asaka}},
  \bibinfo{author}{\bibfnamefont{M.}~\bibnamefont{Laine}}, \bibnamefont{and}
  \bibinfo{author}{\bibfnamefont{M.}~\bibnamefont{Shaposhnikov}},
  \bibinfo{journal}{JHEP} \textbf{\bibinfo{volume}{01}}, \bibinfo{pages}{091}
  (\bibinfo{year}{2007}), \bibinfo{note}{[Erratum: JHEP02,028(2015)]},
  \eprint{hep-ph/0612182}.

\bibitem[{\citenamefont{Merle et~al.}(2014)\citenamefont{Merle, Niro, and
  Schmidt}}]{Merle:2013wta}
\bibinfo{author}{\bibfnamefont{A.}~\bibnamefont{Merle}},
  \bibinfo{author}{\bibfnamefont{V.}~\bibnamefont{Niro}}, \bibnamefont{and}
  \bibinfo{author}{\bibfnamefont{D.}~\bibnamefont{Schmidt}},
  \bibinfo{journal}{JCAP} \textbf{\bibinfo{volume}{1403}}, \bibinfo{pages}{028}
  (\bibinfo{year}{2014}), \eprint{1306.3996}.

\bibitem[{\citenamefont{Merle and Totzauer}(2015)}]{Merle:2015oja}
\bibinfo{author}{\bibfnamefont{A.}~\bibnamefont{Merle}} \bibnamefont{and}
  \bibinfo{author}{\bibfnamefont{M.}~\bibnamefont{Totzauer}},
  \bibinfo{journal}{JCAP} \textbf{\bibinfo{volume}{1506}}, \bibinfo{pages}{011}
  (\bibinfo{year}{2015}), \eprint{1502.01011}.

\bibitem[{\citenamefont{Berlin and Hooper}(2017)}]{Berlin:2016bdv}
\bibinfo{author}{\bibfnamefont{A.}~\bibnamefont{Berlin}} \bibnamefont{and}
  \bibinfo{author}{\bibfnamefont{D.}~\bibnamefont{Hooper}},
  \bibinfo{journal}{Phys. Rev.} \textbf{\bibinfo{volume}{D95}},
  \bibinfo{pages}{075017} (\bibinfo{year}{2017}), \eprint{1610.03849}.

\bibitem[{\citenamefont{Heikinheimo et~al.}(2016)\citenamefont{Heikinheimo,
  Tenkanen, Tuominen, and Vaskonen}}]{Heikinheimo:2016yds}
\bibinfo{author}{\bibfnamefont{M.}~\bibnamefont{Heikinheimo}},
  \bibinfo{author}{\bibfnamefont{T.}~\bibnamefont{Tenkanen}},
  \bibinfo{author}{\bibfnamefont{K.}~\bibnamefont{Tuominen}}, \bibnamefont{and}
  \bibinfo{author}{\bibfnamefont{V.}~\bibnamefont{Vaskonen}},
  \bibinfo{journal}{Phys. Rev.} \textbf{\bibinfo{volume}{D94}},
  \bibinfo{pages}{063506} (\bibinfo{year}{2016}), \eprint{1604.02401}.

\bibitem[{\citenamefont{Bernal and Chu}(2016)}]{Bernal:2015xba}
\bibinfo{author}{\bibfnamefont{N.}~\bibnamefont{Bernal}} \bibnamefont{and}
  \bibinfo{author}{\bibfnamefont{X.}~\bibnamefont{Chu}},
  \bibinfo{journal}{JCAP} \textbf{\bibinfo{volume}{1601}}, \bibinfo{pages}{006}
  (\bibinfo{year}{2016}), \eprint{1510.08527}.

\bibitem[{\citenamefont{Di~Bari et~al.}(2000)\citenamefont{Di~Bari, Lipari, and
  Lusignoli}}]{DiBari:1999ha}
\bibinfo{author}{\bibfnamefont{P.}~\bibnamefont{Di~Bari}},
  \bibinfo{author}{\bibfnamefont{P.}~\bibnamefont{Lipari}}, \bibnamefont{and}
  \bibinfo{author}{\bibfnamefont{M.}~\bibnamefont{Lusignoli}},
  \bibinfo{journal}{Int. J. Mod. Phys.} \textbf{\bibinfo{volume}{A15}},
  \bibinfo{pages}{2289} (\bibinfo{year}{2000}), \eprint{hep-ph/9907548}.

\bibitem[{\citenamefont{Foot and Volkas}(1997)}]{Foot:1996qc}
\bibinfo{author}{\bibfnamefont{R.}~\bibnamefont{Foot}} \bibnamefont{and}
  \bibinfo{author}{\bibfnamefont{R.~R.} \bibnamefont{Volkas}},
  \bibinfo{journal}{Phys. Rev.} \textbf{\bibinfo{volume}{D55}},
  \bibinfo{pages}{5147} (\bibinfo{year}{1997}), \eprint{hep-ph/9610229}.

\bibitem[{\citenamefont{Venumadhav et~al.}(2016)\citenamefont{Venumadhav,
  Cyr-Racine, Abazajian, and Hirata}}]{Venumadhav:2015pla}
\bibinfo{author}{\bibfnamefont{T.}~\bibnamefont{Venumadhav}},
  \bibinfo{author}{\bibfnamefont{F.-Y.} \bibnamefont{Cyr-Racine}},
  \bibinfo{author}{\bibfnamefont{K.~N.} \bibnamefont{Abazajian}},
  \bibnamefont{and} \bibinfo{author}{\bibfnamefont{C.~M.}
  \bibnamefont{Hirata}}, \bibinfo{journal}{Phys. Rev.}
  \textbf{\bibinfo{volume}{D94}}, \bibinfo{pages}{043515}
  (\bibinfo{year}{2016}), \eprint{1507.06655}.

\bibitem[{\citenamefont{Dasgupta and Kopp}(2014)}]{Dasgupta:2013zpn}
\bibinfo{author}{\bibfnamefont{B.}~\bibnamefont{Dasgupta}} \bibnamefont{and}
  \bibinfo{author}{\bibfnamefont{J.}~\bibnamefont{Kopp}},
  \bibinfo{journal}{Phys. Rev. Lett.} \textbf{\bibinfo{volume}{112}},
  \bibinfo{pages}{031803} (\bibinfo{year}{2014}), \eprint{1310.6337}.

\bibitem[{\citenamefont{Bonnivard et~al.}(2015)}]{Bonnivard:2015xpq}
\bibinfo{author}{\bibfnamefont{V.}~\bibnamefont{Bonnivard}}
  \bibnamefont{et~al.}, \bibinfo{journal}{Mon. Not. Roy. Astron. Soc.}
  \textbf{\bibinfo{volume}{453}}, \bibinfo{pages}{849} (\bibinfo{year}{2015}),
  \eprint{1504.02048}.

\bibitem[{\citenamefont{Castorina et~al.}(2012)\citenamefont{Castorina, Franca,
  Lattanzi, Lesgourgues, Mangano, Melchiorri, and Pastor}}]{Castorina:2012md}
\bibinfo{author}{\bibfnamefont{E.}~\bibnamefont{Castorina}},
  \bibinfo{author}{\bibfnamefont{U.}~\bibnamefont{Franca}},
  \bibinfo{author}{\bibfnamefont{M.}~\bibnamefont{Lattanzi}},
  \bibinfo{author}{\bibfnamefont{J.}~\bibnamefont{Lesgourgues}},
  \bibinfo{author}{\bibfnamefont{G.}~\bibnamefont{Mangano}},
  \bibinfo{author}{\bibfnamefont{A.}~\bibnamefont{Melchiorri}},
  \bibnamefont{and} \bibinfo{author}{\bibfnamefont{S.}~\bibnamefont{Pastor}},
  \bibinfo{journal}{Phys. Rev.} \textbf{\bibinfo{volume}{D86}},
  \bibinfo{pages}{023517} (\bibinfo{year}{2012}), \eprint{1204.2510}.

\end{thebibliography}
\end{document}